%% file: arxiv_version.tex
\documentclass[aps,prc,twocolumn,reprint,superscriptaddress,nofootinbib,
tightenlines,preprintnumbers,showkeys,floatfix]{revtex4-2}
 
\usepackage[utf8]{inputenc}
\usepackage[english]{babel}
\usepackage{amssymb,amsthm,amsmath,amstext,amsbsy,amsopn,mathrsfs}
\usepackage{bbm}
\usepackage{nicefrac}
\usepackage{slashed}
\usepackage{graphicx}
\usepackage{hyperref}
\usepackage{leftidx}
\usepackage{environ}
\usepackage{mathtools}
\usepackage{xspace}
\usepackage{array}
\usepackage{xcolor}
\usepackage{isotope}
\usepackage{suffix}
\usepackage{orcidlink}
\usepackage{booktabs,colortbl}
\usepackage[caption=false]{subfig}

\begin{document}

\title{An Efficient Learning Method to Connect Observables}

\author{Hang Yu\,\orcidlink{0000-0001-6860-5960}}
\email{yhang@nucl.ph.tsukuba.ac.jp}
\affiliation{Center for Computational Sciences, University of Tsukuba,
Tsukuba, Ibaraki 305-8577, Japan}

\author{Takayuki Miyagi\,\orcidlink{0000-0002-6529-4164}}
\email{miyagi@nucl.ph.tsukuba.ac.jp}
\affiliation{Center for Computational Sciences, University of Tsukuba,
Tsukuba, Ibaraki 305-8577, Japan}

\begin{abstract}
    Constructing fast and accurate surrogate models is a key ingredient for making robust predictions in many topics.
    We introduce a new model, the Multiparameter Eigenvalue Problem (MEP) emulator. 
    The new method connects emulators and can make predictions directly from observables to observables.
    We present that the MEP emulator can be trained with data from Eigenvector Continuation (EC) and Parametric Matrix Model (PMM) emulators.
    A simple simulation on a one-dimensional lattice confirms the performance of the MEP emulator.
    Using \isotope[28]{O} as an example, we also demonstrate that the predictive probability distribution of the target observables can be straightforwardly obtained through the new emulator.
\end{abstract}

\maketitle

A complete uncertainty quantification on top of large-scale simulations is often considered the most demanding task in computational physics. 
While the computational cost for solving a complicated model multiple times has been addressed by constructing an emulator~\cite{Frame:2017fah,Duguet:2023wuh,Cook:2024toj}, parameters in these models still need to be inferred from the simple $\chi^{2}$ fitting or Bayesian methods~\cite{Bower:2010gal, Vernon:2014gal, Hu:2021trw, Elhatisari:2022zrb, Jiang:2022oba}. 
These procedures not only require a clear understanding and careful implementation of statistical approaches but also combine model biases and parameter uncertainties.
\begin{figure}
    \centering
    \includegraphics[trim={70mm 105mm 45mm 76mm}, clip, width=1\linewidth]{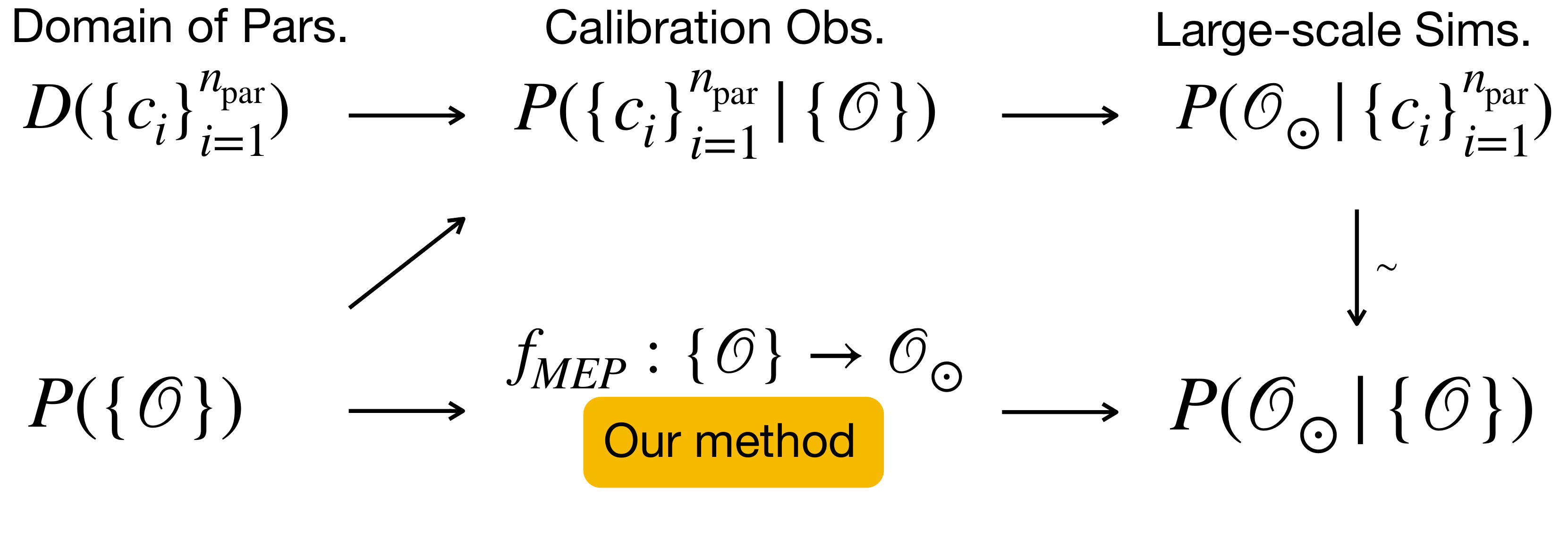}
    \caption{Workflow of our method when compared with existing statistical procedures}
    \label{fig:workflow}
\end{figure}
We propose a new explainable learning model named Multiparameter Eigenvalue Problem (MEP) emulator to both reduce the effort and serve as a first step\footnote{In this work, we focus on the propagation of constraints from input observables to target observables. Systematic uncertainties from model discrepancy and are not addressed in our study.} toward addressing the issues in current statistical methods. The MEP emulator inherits the ideas from and connects the recently developing eigenvector continuation (EC)~\cite{Frame:2017fah} and parametric matrix model (PMM)~\cite{Cook:2024toj}.
Our method allows avoiding parameters by constructing a direct connection between observables and establishes a new bypass to simulate our physical world. 

As the first application,we use it to replace complicated workflow found, for example, in Ref.~\cite{Hu:2021trw, Elhatisari:2022zrb, Jiang:2022oba}. 
Illustrated in the flow chart~Fig.~\ref{fig:workflow}, the conventional workflow starts from a reasonable parameter domain $D(\{c_{i}\}^{n_{\rm par}}_{i=1})$. We then sample the parameter distribution $P(\{c_{i}\}^{n_{\rm par}}_{i=1}|\{\mathcal{O}\})$ calibrated with a set of known observables $\{\mathcal{O}\}$. Finally, the conditional posterior distribution $P(\mathcal{O}_{\odot}| \{\mathcal{O}\})$ for a target observable $\mathcal{O}_{\odot}$ is obtained through the marginalization.
The underlying physics is grounded in the low-energy nuclear \emph{ab initio} theory, which is inherently computationally demanding and multiparametric~\cite{Hergert2020, Ekstrom2023}. 
Many parameters called low-energy constants (LECs) can appear at the same order in the expansion of the underlying effective field theory~\cite{Weinberg:1978kz,Epelbaum2009,Machleidt2011,Hammer:2019poc}.
These parameters cannot be directly measured, and the determination of LECs already can experience difficulties at the two-body level and requires a special strategy~\cite{Peng:2024aiz}.
By introducing the MEP emulator, we construct a function from $\{\mathcal{O}\}$ to $\mathcal{O}_{\odot}$, $f_{\rm MEP}: \{\mathcal{O}\} \to \mathcal{O}_{\odot}$, and directly obtain the conditional distribution $P(\mathcal O_\odot| \{\mathcal O\})$.
This avoids the non-measurable parameters, provides a direct way to leverage a strong correlation via physical observables, and hence, unifies efficient emulation and uncertainty quantification into one simple emulator\footnote{There is still marginal dependence on the choice of parameters to generate training samples. However, we expect this dependence to be suppressed with the increasing of number of samples.}.
Despite our focus on the uncertainty quantification of nuclear physics, our emulator has its roots in the Ritz approximation method, enabling possible generalization to problems ranging from quantum chemistry to engineering physics~\cite{Plestenjak:2015spe,Kiefer:2023com} that need multiple parameters. 
In this work, we focus on the observables obtained from Hamiltonian.

Before directly discussing the structure of the MEP emulator, we would like to begin with a simple idea.
In many situations, we usually encounter the following Hamiltonian equation with the affine form:
{\color{blue} }
\begin{equation}
    \left(H_0 + \sum_{i=1}^{n_{\rm par}} c_{i} H_{i} - E N\right) \mathbf{y} = 0\,,
    \label{eq:emu}
\end{equation}
with a nontrivial vector $\mathbf{y}$. 
Here, $H_{0}$ is a parameter-free part of the Hamiltonian, $c_{i}$ is $i$-th parameter, $H_{i}$ is a part of the Hamiltonian depending on $c_{i}$, $E$ is the eigenvalue of the Hamiltonian, and $N$ is the norm kernel.
A similar structure can be found in reduced order models such as EC methods~\cite{Frame:2017fah}, where the matrix elements of $H_{0}$, $H_{i}$, and $N$ are computed with the training vectors at given $\{c_{i}\}^{n_{\rm par}}_{i=1}$.
It is less discussed, but writing the equation in the above way, $E$ can be regarded as an input, as in scattering problems. Then, for example, $-c_{n_{\rm par}}$ can be regarded as the eigenvalue of the equation after multiplying the inverse $H^{-1}_{n_{\rm par}}$ (assuming it exists):
\begin{equation}
  H_{n_{\rm par}}^{-1}   \left(H_0 + \sum_{i=1}^{n_{\rm par}-1} c_{i} H_{i} - E N\right)\mathbf y  + c_{n_{\rm par}} \mathbf y =0\,.
  \label{eq:inversion}
\end{equation}
This input-output exchange hints our theoretical idea behind this work: we can make predictions without directly addressing the parameters in the Hamiltonian.
This simple scheme allows us to predict other quantities using $E$ as the input if only $c_{n_{\rm par}}$ is varied. 
But in reality,
for example, in nucleon-nucleon interaction from chiral EFT in nuclear physics~\cite{Epelbaum2009, Machleidt2011}, we generally expect that 10 - 30 parameters should be varied simultaneously to make meaningful predictions. 
Therefore,  we couple a same number of Eq.~\eqref{eq:emu} through a new mathematical technique -- our MEP emulator.
The number of LECs will continue to grow with our knowledge about three-nucleon interaction~\cite{Girlanda:2011fh,Epelbaum:2014sea}.

In the following, we consider the $m (> n_{\rm par})$ Hamiltonian equations that take $E^{[1]}, \ldots, E^{[n_{\rm par}]}$ as inputs (to trade $n_{\rm par}$ parameters $c_i$) and $E^{[n_{\rm par}+1]}, \ldots, E^{[m]}$ as outputs:
\begin{equation}
\left( H_{0}^{[j]} + \sum_{i=1}^{n_{\rm par}} c_{i} H_{i}^{[j]}    - E^{[j]} N^{[j]} \right) \mathbf{y}^{[j]} = 0\,,~~~ 1\leq j\leq m\,.
\label{eq:Main}
\end{equation}
Eq.~\eqref{eq:Main} resembles the multiparameter eigenvalue problem (MEP) \cite{atkinson:1968mus, Atkinson:2010mul}, i.e.,
\begin{equation}
\left(O_{j} + \sum_{i=1}^{m} \alpha_i A_{ij} \right) \mathbf{y}_j = 0\,,~~~ 1\leq j\leq m\,.
\label{eq:MEP}
\end{equation}
Here, $O_{j}$ and $A_{ij}$ are square matrices, and $\alpha_{i}$ is a parameter of the \textit{generalized} eigenvalue problem, hence it is referred to as a \textit{multiparameter eigenvalue}. The solution is obtained by solving the  generalized eigenvalue problem:
\begin{equation}
    (K_i - \alpha_i K_0)\mathbf y_\otimes =0\,,
    \label{eq:GEM}
\end{equation}
with the generalized (Kronecker) determinants $K_i$ and $K_0$.
\begin{equation}
    K_i =\begin{vmatrix}
A_{11} &\cdots &A_{(i-1)1} &O_{1}& A_{{(i+1)1}}& \cdots & A_{m1} \\
\vdots & \vdots & \vdots &\vdots &\vdots & \vdots & \vdots \\
A_{1m}  & \cdots&A_{(i-1)m} &O_{m}& A_{{(i+1)m}}& \cdots & A_{mm}
\end{vmatrix}_{\otimes},
\label{eq:Ki}
\end{equation}
and 
\begin{equation}
    K_0 =\begin{vmatrix}
A_{11}  &\cdots & A_{m1} \\
\vdots  & \ddots & \vdots \\
A_{1m} & \cdots& A_{mm}
\end{vmatrix}_{\otimes}.
\label{eq:Ko}
\end{equation}
This determinant is computed by replacing the product operation with the Kronecker product, defined in Eq.~\eqref{eq:krondet} in Appendix~\ref{app:inner_product}.
Different interpretations of Eq.~\eqref{eq:Main} into Eq.~\eqref{eq:MEP} compose a superset of Hamiltonian multiparameter eigenvalue problems. 
For example, Eq.~\eqref{eq:Main} can be found with $O_{j} = H_{0}^{[j]} +  \sum_{i=1}^{n_{\rm par}} c_{i} H_{i}^{[j]}$, $A_{ij} = \delta_{ij}N^{[j]}$, and  $\alpha_{i} = E^{[i]}$. 
This choice yields a set of decoupled MEP equations, i.e., $A_{ij} \propto \delta_{ij} $.
Solving this system precisely corresponds to computing energy structures from a given parameter set $\{c_{i}\}_{i=1,\ldots, n_{\rm par}}$.
On the other hand, when one requires to use $\{E^{[j]}\}_{j=1, \ldots, n_{\rm par}}$  as an input instead of $c_{i}$ to predict $\{E^{[j]}\}_{j=n_{\rm par}+1,\ldots, m}$, the following can be applied: 
\begin{equation}
\begin{aligned}
O_{j} &= \left\{
\begin{array}{rr}
H_{0}^{[j]} - E^{[j]}N^{[j]} & 1 \leq j \leq n_{\rm par} \\
H_{0}^{[j]} & n_{\rm par} < j \leq m
\end{array}
\right. \, ,\\
A_{ij} &= \left\{
\begin{array}{rr}
H_{i}^{[j]} & 1 \leq j \leq n_{\rm par} \\
-\delta_{ij}N^{[i]} & n_{\rm par} < j \leq m
\end{array}
\right. \, ,\\
\alpha_{i} &= \left\{
\begin{array}{rr}
c_{i}& 1 \leq i \leq n_{\rm par} \\
E^{[i]} & n_{\rm par} < i \leq m
\end{array}
\right. \,.
\end{aligned}
\end{equation}
With this rearrangement, $A_{ij}$ for $i\neq j$ does not vanish anymore, and MEP becomes numerically expensive.
Assuming that each equation in Eq.~\eqref{eq:MEP} has $M\times M$ dimension, the dimension of $K_{1}$ and $K_{i}$ is $M^m \times M^m$, and thus the size of the whole problem is not small as we have $m = 10 \sim 30$ in a typical nuclear physics application, even typical reduced basis matrix can achieve good accuracy with $M\sim 10$  .
Problems with such large dimension are unsolvable on modern supercomputers.
Even worse, we have not found a way to select the desired eigenvalue $E^{[j]} \, (n_{\rm par} < j \leq m)$ from the full set of $M^{m}$ eigenvalues.
This point will be emphasized with a simple toy model, and a direct application of the known MEP to our problems is not suitable.

To overcome these issues, we consider  reducing the problem size based on the projection emulators~\cite{Frame:2017fah}.
Suppose we have a set of $n_{\rm{train}}$ training parameters $S_{\rm{train}} = \{\mathbf{c}_{k}| \mathbf{c}_{k} = (c_{1,k},c_{2,k}\ldots, c_{n_{\rm{par}},k}),\; k = 1,\ldots,n_{\rm{train}} \} $. 
Using the $k$-th training parameter set $\mathbf{c}_{k}$, we have an eigenvector $\mathbf y_{j,k} $ 
for the $j$-th equation in Eq.~\eqref{eq:MEP}. 
The product of eigenvectors in Eq.~\eqref{eq:GEM} is then $\mathbf y_{\otimes,k} = \bigotimes_{j=1}^m \mathbf y_{j,k} $. 
The Kronecker determinant reduces to the usual matrix determinant after taking the inner product with $k$-th and $l$-th eigenvectors:
\begin{equation}
    (\mathbf y_{\otimes,k})^* \cdot K_0 \cdot\mathbf y_{\otimes,l} =
     \begin{vmatrix}
(A_{11})_{kl} & \cdots &(A_{m1})_{kl}   \\
\vdots & \ddots & \vdots \\
(A_{1m})_{kl} & \cdots& (A_{mm})_{kl} 
\end{vmatrix},
\label{eq:inneremu}
\end{equation}
generally applies to Kronecker determinants. 
Further details can be found in Appendix~\ref{app:inner_product}.
The object $(A_{ij})_{kl}\equiv(\mathbf{y}_{j,k})^{*} \cdot A_{ij} \cdot \mathbf{ y}_{j,l}$, coincides with the $kl$-th element  of subspace projected $A_{ij}$, and can be retrieved directly from original Hamiltonians. 
These indices may appear counterintuitive, see Eq.~\eqref{eq:det_prod} for details.
Once we know how to compute these inner products, we can explicitly write our MEP emulator for Eq.~\eqref{eq:GEM} from the subspace projection perspective. This projection works even if the Hamiltonian equations~\eqref{eq:Main} are given by emulators.
We denote the reduced order matrices for $K_0$ and $K_i$ by $\mathcal K_0$ and $\mathcal K_i$, with their $kl$-th entry computed from Eq.~\eqref{eq:inneremu},
\begin{align}
    (\mathcal K _i)_{kl} & = (\mathbf y_{\otimes,k})^* \cdot K_i \cdot\mathbf y_{\otimes,l},\\
    (\mathcal K _0)_{kl} & = (\mathbf y_{\otimes,k})^* \cdot K_0 \cdot\mathbf y_{\otimes,l}.
\end{align}
Solving 
\begin{equation}
(\mathcal K _i - \alpha_i \mathcal K _0)\mathbf{y} =0,
\label{eq:MEPemu}
\end{equation}
yields the desired emulator $f_{\rm MEP}$ illustrated in Fig.~\ref{fig:workflow}, a direct mapping from observables to observables  $\alpha_i$.

Besides the connection with the EC emulators, we notice that our reduced order model Eq.~\eqref{eq:MEPemu} preserves the nice affine form found in the original problem Eq.~\eqref{eq:Main}.
Note that the inputs of our MEP problem always enter the $O_{j}$ matrices, they are in the same column in the determinant Eq.~\eqref{eq:Ki}.
Calculating the inner product Eq.~\eqref{eq:inneremu} will always carry at maximum the same order of $\{E^{[j]}\}_{j=1, \ldots, n_{\rm par}}$ as it will appear in the original equation. 
Hence, we can always find 
\begin{multline}
     (\mathcal K _i - \alpha_i \mathcal K _0)\mathbf{y} 
     =  (\mathcal K _{i0} + E^{[1]} \mathcal K _{i1} +  \ldots \\ + E^{[n_{\rm par}]} \mathcal K _{in_{\rm par}} - \alpha_i \mathcal K _0)\mathbf{y} = 0\,.
     \label{eq:MEPMM}
\end{multline}
This observation generalizes our application from EC to affine PMM methods~\cite{Cook:2024toj} and can connect affine PMM and EC emulators. 
We shall focus on a demonstration of the MEP emulator in this work and defer the details of hybrid applications to the future.

\begin{figure}[t]
    \includegraphics[trim={0 15mm 5mm 0mm},clip,width=1\linewidth]{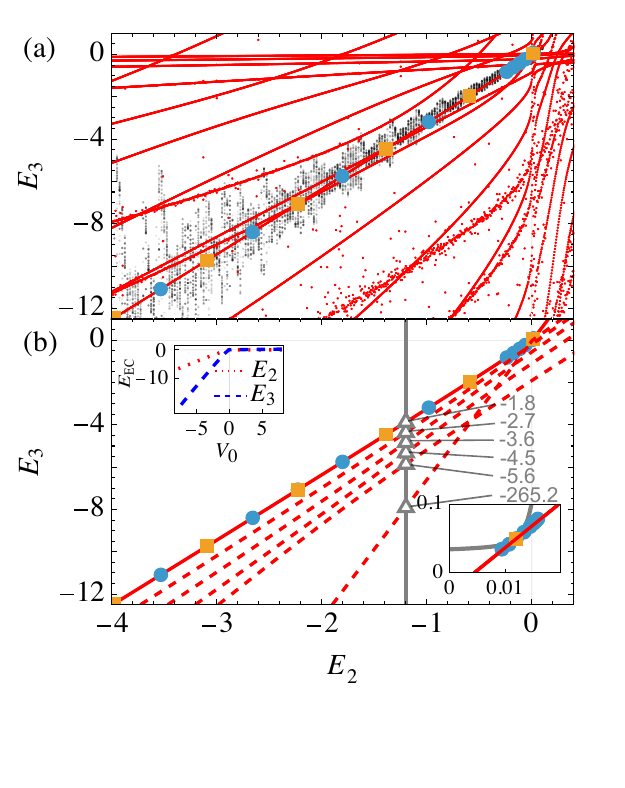}
    \caption{Comparing methods mapping from energy  of a two-body system $E_{2}$ to energy of a three-body system $E_{3}$. This figure demonstrates the robustness of our MEP emulator compared to direct MEP calculations and conventional inversions. 
    The energies are computed with a simple Hamiltonian in the one-dimensional lattice, see text for the details.
    The training points for the EC emulator are represented by orange squares; blue dots represent validation from the original problem Eq.~\eqref{eq:Main}.
    In panel (a), the red dots and lines are computed by solving Eq.~\eqref{eq:GEM} for EC. The gray dots are obtained from the spectra of the original problem.
    Panel (b) shows the energies computed from $f_{\rm MEP}$, Eq.~\eqref{eq:MEPemu}.
    The solid (dashed) red curve corresponds to ground-states $E_{2}$ and $E_{3}$ (excited-states $E_{2}$ and $E_{3}$).
    The triangles emphasize the corresponding $V_{0}$ for the given $E_{2}$ and  $E_{3}$ .
    The ground-state (emulated by EC) $E_{2}$ and $E_{3}$ as a function of $V_{0}$ are show in the top inset.
    As a perfomance metric, the gray curve in the zoomed-in plot on the bottom is from inverting EC emulators according to Eq.~\eqref{eq:inversion} (see text). }
    \label{fig:MEP}
\end{figure}
Here, we show a proof-of-principle application of the MEP emulator with a simple one-dimensional lattice toy model based on Ref.~\cite{Konig:2017krd}.
In the model, we set all the quantities as dimensionless and particle masses to unity, and the Hamiltonian is defined as $H = \sum_{i}^{N} T_{i} + \sum_{i<j}^{N}V_{ij}$ with the particle number $N$, one-body kinetic term $T_{i}$, and two-body interaction $V_{ij}$.
The numerical calculations are performed within the space with lattice size $L=30$ and spacing $a_{\rm latt}=0.25$.
For $T_{i}$, the $\mathcal O (a_{\text{latt}}^3)$ improved lattice kinetic energy operator \cite{Lahde:2019npb} is used.
Also, the two-body contact interaction smeared with the Gaussian is employed; $V_{ij} = V_0 \exp(-r_{ij}^2/R^2)$, where $r_{ij}$ is the distance between the particles $i$ and $j$.
Throughout this work, we use $R=2$.
With this one-dimensional lattice model, we compute the ground states for $N=2$ and $3$ with various $V_{0}$ to investigate the performance of the MEP emulator.
To do so, we first construct the EC emulators for two- and three-body systems with the training samples at $V_0 = 0.5$, $-1$, $-2$, $-3$, $-4$, and $-5$.

In Figure~\ref{fig:MEP} (a), we emulate the two- and three-body ground-state energies $E_{2}$ and $E_{3}$.
Here, $E_{3}$ is obtained by solving the original MEP equation~\eqref{eq:GEM} as a function of $E_{2}$ instead of $V_{0}$.
As seen in the figure, $M^{m}=36$ possible eigenvalues show complicated level crossings and branch switchings everywhere across the training domain, while the original lattice MEP equations have roughly $10^6$ eigenvalues in total\footnote{We only approximately (with larger lattice spacing $a_{\text{latt}} = 1$) solve the MEP Eq.~\eqref{eq:GEM} for the original lattice models, finding only 100 eigenvalues (gray scattered points in Fig.~\ref{fig:MEP} (a)) around the exact solution. }. 
These crossings prevent a simple and general numerical scheme to extract the desired three-body ground-state energy for our problem. 
Furthermore, in EC emulators, training vectors usually are highly correlated, consequently, their normal matrices are not well-conditioned~\cite{Duguet:2023wuh}. 
These not well-conditioned matrices hinder the accuracy of numerical operations in solving the MEP equations. 
We see many isolated dots in Fig.~\ref{fig:MEP} (a) that are obviously numerical artifacts. These two major issues forbid any meaningful interpretation of the results.

We proceed to check our MEP emulator Eq.~\eqref{eq:MEPemu} in Fig.~\ref{fig:MEP} (b).
The MEP emulator $f_{\rm MEP}$ is constructed with the ground-state wave functions.
With the $M\times M =6 \times 6$ matrices, we can retrieve all our training data on the same branch (solid curve) that does not suffer from any switching.
We also do not notice any numerical artifacts observed in Fig.~\ref{fig:MEP} (a).
With the MEP emulator, it is possible to select the ground-state $E_{3}$ by comparing $V_{0}$ computed from Eq.~\eqref{eq:MEPemu}.
At a fixed $E_{2}$, one can find different $V_{0}$ as shown by the triangles in the figure, and the branches indicated by the dashed curves correspond to excited-state $E_{2}$ obtained with a more attractive interaction.
The strength $V_{0}$ can be efficiently retrieved because it shares the same eigenvector with the corresponding $E_3$. 
This scheme can be easily extended to general applications.

We also compare our emulator to the direct inversion Eq.~\eqref{eq:inversion}, when
the simple inversion is possible with a single varied parameter $V_0$. 
From the inset of the figure, we already notice a significant improvement in the scattering region.
This is because the boundary condition heavily suppresses the $V_0$ dependence of discrete scattering eigenvalues~\cite{Luscher:1990ux}.
Inverting this almost flat dependence will significantly amplify any numerical difference in Eq.~\eqref{eq:inversion} when emulating the energy levels.
Only our new emulator can hold up in this simple toy test. We conclude that our MEP emulator is the only existing solution in some highly non-linear regions.
\begin{figure}[t!]
    \centering
    \includegraphics[trim={2mm 2mm 12mm 12mm}, clip, width=1\linewidth]{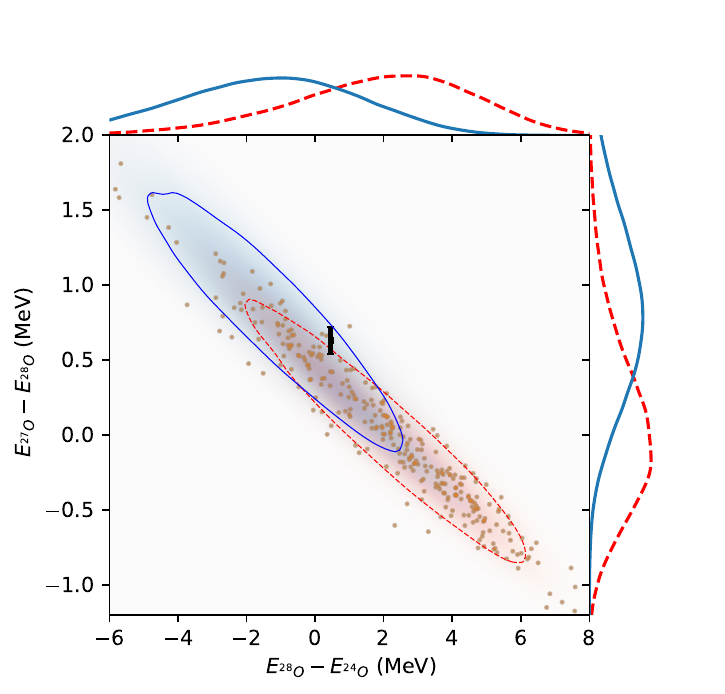}
    \caption{Joint distribution of $P(\mathcal{O}_{\odot}| \{\mathcal{O}\})$, where $\mathcal{O}_{\odot}$ are \isotope[28]{O} energy differences  with respect to \isotope[27]{O} and \isotope[24]{O}. The red shade and red dashed lines are our emulator prediction using $10^5$ samples from averaging 10 resampling sets with input $P(\{\mathcal O\})$  constructed directly from~\cite{Jiang:2022oba}. We then calibrate our joint density (blue shades and blue solid lines) with input  $P_C(\{\mathcal O\})$ from additional constraints on ground states of $A = 16, 24$ (see text).
    The small orange dots are the data points used in constructing the emulators. We do not have any input from LECs. Circles indicate $68$\% confidence intervals (CIs). Black error bars are experimental values taken from Ref.~\cite{Kondo:2023lty}.}
    \label{fig:o28}
\end{figure}

For further validation, we consider a realistic problem from a data-driven aspect.
In the following, we present the results using the valence-space in-medium similarity renormalization group (VS-IMSRG)~\cite{Stroberg:2019mxo, Hergert:2020bxy} to explore the energy of the newly discovered \isotope[28]{O}~\cite{Kondo:2023lty}.
To this end, we numerically train the MEP emulator~\eqref{eq:MEPMM} using the PMM approach~\cite{Cook:2024toj}. 
This approach allows us to directly find the matrices in Eq.~\eqref{eq:MEPMM} (see Appendix~\ref{app:PMM} for more details).
We produce the training data using 300 non-implausible (NI) samples \cite{Jiang:2022oba} of Delta-full chiral EFT interactions at the next-to-next-to-leading order~\cite{Ekstrom:2017koy} with momentum cutoff at $\Lambda = 394$ MeV. 
The VS-IMSRG calculations are done within the 13 major-shell harmonic-oscillator (HO) space with the frequency parameter $\hbar\omega = 16$ MeV.
Another truncation needs to be introduced for the three-nucleon matrix elements and is defined as $E_{\rm 3max}$ with the sum of the three-nucleon HO quanta. 
In this work, the sufficiently large $E_{\rm 3max}=18$ is used.
With some calculations, we expect that the basis truncation error is around $0.2$ MeV for \isotope[24]{O}. 
We train two MEP emulators with the $12 \times 12$ dimension matrices using observable inputs to predict two observables $\mathcal O_{\odot}$: the ground-state energy differences of \isotope[27]{O}, \isotope[28]{O} and \isotope[28]{O}, \isotope[24]{O}.
Our inputs include ground-state energies of \isotope[2]{H}, \isotope[3]{H}, \isotope[4]{He}, \isotope[6]{Li}, \isotope[16]{O} and \isotope[24]{O}, as well as proton-neutron scattering phase shifts in the NI samples at lab energies $E = 5, 50$ MeV.
To make our system not overdetermined, we remove one phase shift at $E = 50$ MeV in $^{3}P_{2}$ channel so that the total number of input observables equals the number of LECs. 
These observables are our input throughout the current study, denoted as $\{\mathcal O\}$ in Fig.~\ref{fig:workflow}.
We do not notice any differences in adding more (highly correlated) phase shifts and making this system overdetermined, while removing ground state energies often leads to failed training.
PMM methods will work once inputs are adapted to observables: in data-driven MEP emulators, we need not explicitly construct an emulator to access observables such as phase shift. Instead, the MEP trick Eq.~\eqref{eq:MEPMM} was applied, and we compute the reduced order matrices  $\mathcal K_{ij}$ using PMM method directly. More details can be found in Appendix~\ref{app:phase_shift} and~\ref{app:PMM}.

Our MEP emulator constructs an efficient function $f_{\rm{MEP}}$ directly mapping from $\{\mathcal O\}$ to $\mathcal O_{\odot}$. With this deterministic $f_{\rm{MEP}}$ , we can explore the energy properties of \isotope[28]{O} by sampling a proper distribution of the input $\{\mathcal O\}$. This $f_{\rm{MEP}}$ corresponds to solving the generalized eigenvalue problems of $12 \times 12$ matrices, enabling us to compute millions of samples in a few minutes. We can then perform frequentist uncertainty estimation, which allows us a direct comparison without assuming priors with the recent experiment for \isotope[28]{O}~\cite{Kondo:2023lty}.
In Fig.~\ref{fig:o28}, the resulting conditional distributions $P(\mathcal{O}_{\odot}| \{\mathcal{O}\})$ are illustrated.
To choose an input condition $P(\{\mathcal O\})$, we begin with the multivariate Gaussian distribution by computing the covariance matrix from equally weighted 8188 NI sample data from Ref.~\cite{Jiang:2022oba}. This $P(\{\mathcal O\})$ is the simplest choice to cover the whole NI space.
Ref.~\cite{Kondo:2023lty} uses $A = 16 - 24$ observables as calibration observables in their coupled-cluster calculations. We also noticed that our simple $P(\{\mathcal O\})$ often produce biased trial samples for \isotope[16]{O} and \isotope[24]{O} ground states.
To explore the impact of calibrations, we replace $P(\{\mathcal O\})$ with our calibrated distribution  $P_C(\{\mathcal O\})$  through conditioning $P(\{\mathcal O\})$ with marginal distributions centered at the experimental values of $E_{\isotope[16]{O}} = -127 \pm 2$~MeV and $E_{\isotope[24]{O}} = -168 \pm 3$~MeV.
Note that standard deviations are \textit{ad hoc} estimations based on model uncertainties~\cite{Kondo:2023lty}.
The corresponding probability distribution is given by the solid line in the figure.
Compared with the literature \cite{Kondo:2023lty}, the results before the calibration are already consistent with experimental results, with our 68\% CI overlapping with the experimental CI. 
After the calibration, it is observed that our joint distribution moves closer to the experimental value, and the new 68\% CI overlaps with the experimental CI better. 
We note that our distribution consistently overlaps with the prediction by the coupled-cluster calculations~\cite{Kondo:2023lty}.
The clear improvement in Fig.~\ref{fig:o28} indicates the need for additional calibrations based on few-body observables as well. 
This example shows that our MEP emulator can be used to efficiently explore constraints given by observables, produce consistent predictions, and generate guidance on future improvements.


To conclude, we introduce a novel deterministic MEP emulator from the theory of the multiparameter eigenvalue problem to connect different model predictions.
This new emulator is applicable to both model- and data-driven approaches.
It is observed that the direct application of the multiparameter eigenvalue problem to a nuclear physics study is impossible without reduced basis methods. 
From the model-driven perspective, we thoroughly benchmark our MEP emulator and show its advantage in non-linear domains. 
As an application to a realistic problem, we discuss the probability distribution for the energy properties of the newly discovered \isotope[28]{O}.
The MEP emulator can be directly constructed from existing data tables and allows us to improve the workflow to understand and predict properties of atomic nuclei and to efficiently explore correlations between observables obtained from Hamiltonian.
Since the proposed procedure is general, one can expect potential adaptations to many different applications, including Multi-fidelity inputs~\cite{Belley:2023lec}, the volume-dependence perspective~\cite{Luscher:1990ux}, and exploring time-dependent problems \cite{Atkinson:2010mul}. 


\begin{acknowledgments}
We thank Dean Lee for enlightening discussions and Nobuo Hinohara for carefully reading the
manuscripts and many comments. This work is in part supported by JST ERATO Grant No. JPMJER2304, Japan. This work is also in part supported by the Multidisciplinary Cooperative Research Program in CCS, University of Tsukuba.
\end{acknowledgments}

\bibliography{refs}

\appendix
\input{endmatter}

\end{document}

%% file: endmatter.tex
\section{Additoinal comments \label{app:phase_shift}}
We emphasize that the emulator constructed in this work represents a distinct paradigm compared to established methods such as Eigenvector Continuation (EC)~\cite{Frame:2017fah} and Parametric Matrix Models (PMM)~\cite{Cook:2024toj}. While EC and PMM primarily operate as mappings from Hamiltonian parameters to physical quantities—with EC specifically relying on wavefunction snapshots to build a projection subspace—our method (MEP) utilizes physical observables directly as inputs. This distinction allows for emulator construction even when the underlying microscopic wavefunction is unavailable or computationally prohibitive. A comparison of these approaches is summarized in Table~\ref{tab:compare}.
\begin{table}[hbtp]
    \centering
    \begin{tabular}{ccc}
\textbf{Method} & \textbf{Input} & \textbf{w.f?} \\\hline
        EC~\cite{Frame:2017fah}  & Hamiltonian Parameters & Yes \\
        PMM~\cite{Cook:2024toj} & Hamiltonian Parameters & No \\
        \textbf{MEP (This work)} & \textbf{Observables}   & \textbf{Optional} \\
    \end{tabular}
    \caption{Comparison of emulator methodologies. We distinguish methods by their input variables (Hamiltonian parameters vs. physical observables) and their reliance on access to the microscopic wavefunction (w.f.?) during the training or construction phase.}
    \label{tab:compare}
\end{table}

For MEP, we note that $E^{[i]}$ in Eq.~\eqref{eq:Main} is not necessarily an energy. A simple and notable example is the phase shift. To that end, we can solve the system:
\begin{equation}
    \left[H_0 + c_1 H_1  +\ldots + c_n H_n  + V_{\rm wall}(L) -\frac{p^2}{2\mu} \right] \mathbf{y} = 0\,,
\label{eq:cot}
\end{equation}
using the spherical wall method \cite{Borasoy:2007vy} with $p>0$ momenta of this system. Quantization condition gives $ p L = -\delta + n \pi $, where integer $n$ can be obtained from the system we are studying. For instance, $n =0$ for ground states in an  unbounded system by the Levinson's theorem. We can therefore translate phase shift $\delta$ to sphere radii $L$ or momenta $p$ as an input, whichever is convenient. It is worth noting that this is a typical example where we do not require the input/output to be linear. 
This procedure has been numerically verified from training several PMM emulators for phase shift using the data from Ref.~\cite{Jiang:2022oba}. We will discuss further details in a forthcoming publication

To incorporate phase-shift into MEP emulators, we note Eq.~\eqref{eq:cot} can be solved together with Eq.~\eqref{eq:Main}, resulting a general form of Eq.~\eqref{eq:MEPMM}. 
Since we used the PMM method to numerically determine the $\mathcal {K}_{ij}$, in this work, an explicit phase shift emulator construction is not necessary.

\section{The inner product formula \label{app:inner_product}}
Here, we derive the inner product in Eq.~\eqref{eq:inneremu}.
Without losing generality, one can compute the inner product of $\mathbf{u}_\otimes^{*} \cdot K \cdot \mathbf{v}_\otimes$ in Eq.~\eqref{eq:GEM}.
The definition of the Kronecker determinant $K$ is 
\begin{multline}
   K=  
   \begin{vmatrix}
G_{11}  &\cdots & G_{m1}  \\
\vdots& \ddots  & \vdots \\ 
G_{1m} & \cdots& G_{mm}
\end{vmatrix}_{\otimes} \\
=  \sum_{\sigma\in{S_m}} {\mathrm{sgn}(\sigma) } G_{\sigma(1)1}\otimes G_{\sigma(2)2}\ldots\otimes G_{\sigma(m)m}\\
\equiv \sum_{\sigma\in{S_m}} {\mathrm{sgn}(\sigma) } K_\sigma \,.
\label{eq:krondet}
\end{multline}
Then, we can write 
\begin{equation}
    \mathbf{u}_\otimes^* \cdot K \cdot \mathbf{v}_\otimes = 
    \sum_{\sigma\in{S_m}} {\mathrm{sgn}(\sigma) } \mathbf{u}_\otimes^* K_\sigma \mathbf{v}_\otimes \,.
\end{equation}
$\mathbf{u}_\otimes$ and $\mathbf{v}_\otimes$ are vectors belonging to the vector space of the corresponding linear operator $K$. Therefore, we can write 
\begin{equation}
    \mathbf{u}_\otimes = \mathbf{u}_1\otimes \mathbf{u}_2 \otimes \ldots \otimes \mathbf{u}_m\,,\,
\end{equation}
and 
\begin{equation}
    \mathbf{v}_\otimes = \mathbf{v}_1\otimes \mathbf{v}_2  \otimes \ldots \otimes \mathbf{v}_m\,.
\end{equation}
Then, the product can be written as
\begin{multline}
     \mathbf{u}_\otimes^* \cdot K_\sigma \cdot \mathbf{v}_\otimes  =
     (\mathbf{u}_1^*\otimes \mathbf{u}_2^* \otimes \ldots \otimes \mathbf{u}_m^*)
      \\ \cdot (G_{\sigma(1)1}\otimes G_{\sigma(2)2}\otimes \ldots\otimes G_{\sigma(m)m}) \\\cdot
     (\mathbf{v}_1\otimes \mathbf{v}_2  \otimes \ldots \otimes \mathbf{v}_m).
\end{multline}
As the Kronecker product satisfies $(A\otimes B) \cdot (C\otimes D)\cdot ( E\otimes F) = [(AC)\otimes(BD)]\cdot(E\otimes F) = (ACE)\otimes(BDF)$, one can find
\begin{multline}
     \mathbf{u}_\otimes^* \cdot K_\sigma \cdot \mathbf{v}_\otimes  =
     (\mathbf{u}_1^* G_{\sigma(1)1} \mathbf{v}_1 ) \otimes \ldots  \otimes (\mathbf{u}_m^* G_{\sigma(m)m} \mathbf{v}_m ) 
\end{multline}
The product $\mathbf{u}_i^* G_{i\sigma(i)} \mathbf{v}_i$ becomes a scalar, and thus
\begin{multline}
     \mathbf{u}_\otimes^* \cdot K_\sigma \cdot \mathbf{v}_\otimes  =\\
    (\mathbf{u}_1^* G_{\sigma(1)1} \mathbf{v}_1 ) \ldots  (\mathbf{u}_m^* G_{\sigma(m)m} \mathbf{v}_m ) 
\end{multline}
is also a scalar.
Substituting the result for Eq.~\eqref{eq:krondet}, we obtain
\begin{multline}
     \mathbf{u}_\otimes^* \cdot K \cdot \mathbf{v}_\otimes = \sum_{\sigma\in{S_m}}{\mathrm{sgn}(\sigma) } \\
     \cdot(\mathbf{u}_1^* G_{\sigma(1)1} \mathbf{v}_1 ) \cdot (\mathbf{u}_2^* G_{\sigma(2)2} \mathbf{v}_2 ) \ldots  \cdot (\mathbf{u}_m^* G_{\sigma(m)m} \mathbf{v}_m ),
\end{multline}
which coincides with the definition of the matrix determinant:
\begin{multline}
\label{eq:det_prod}
     \mathbf{u}_\otimes^* \cdot K \cdot \mathbf{v}_\otimes =  \begin{vmatrix}
u_1^* G_{11} v_1   &\cdots & u_1^* G_{m1} v_1   \\
\vdots& \ddots  & \vdots \\
u_m^* G_{1m} v_m & \cdots& u_m^* G_{mm} v_m
\end{vmatrix}.
\end{multline}

\section{PMM for VS-IMSRG \label{app:PMM}}
\begin{figure*}[hbtp]
\centering
\subfloat[PMM  for $^{16}${O}]{
        \includegraphics[width=0.2453\textwidth]{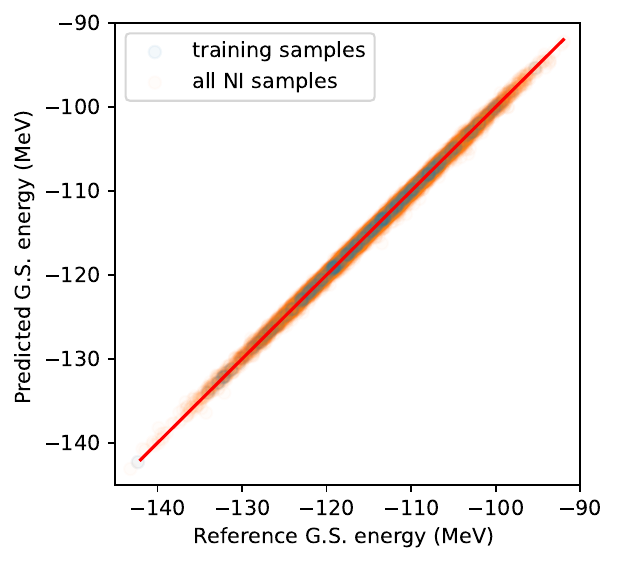}
        \label{fig:em_PMM}
    }
\subfloat[PMM  for  $^{24}${O}]{
        \includegraphics[width=0.2365\textwidth]{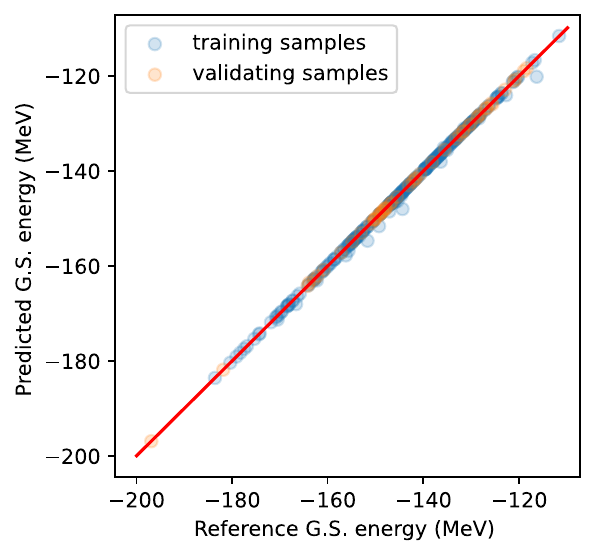}
        \label{fig:em_24O}
    }
\subfloat[MEP  for   $^{24}${O}]{
        \includegraphics[width=0.2365\textwidth]{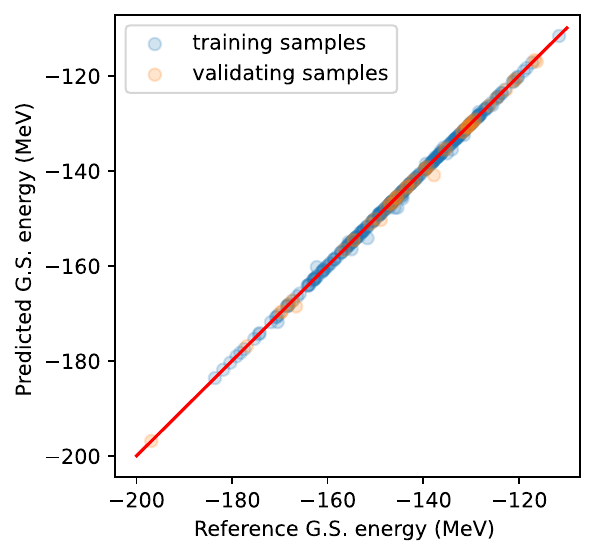}
        \label{fig:em_MEPMM}
}
\subfloat[MEP for $E_{\isotope[28]{O}}-E_{\isotope[24]{O}}$ ]{
        \includegraphics[width=0.234\textwidth]{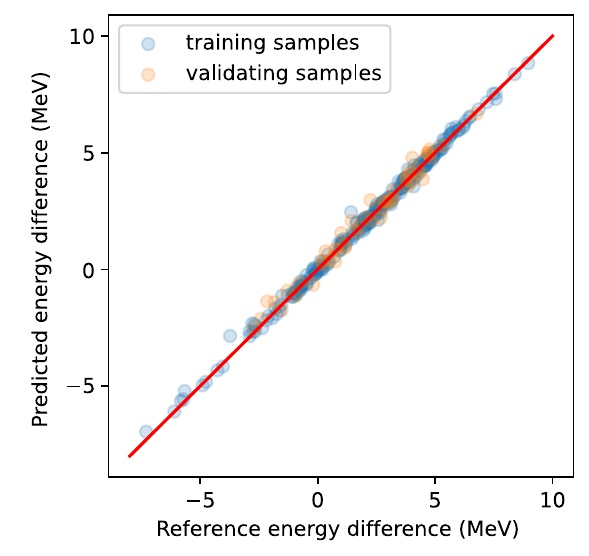}
        \label{fig:em_MEPdiff}
    }
    \caption{Performance of emulators. Panels (a) show the PMM performance for the 8188 samples of ground-state energy for  \isotope[16]{O}. Panel (b) and (c) shows the performance for our VS-IMSRG emulators on \isotope[24]{O} ground-states energy where the emulators take the 17 LECs and observables as inputs, respectively. Panel (d) is for our MEP emulator targeting the ground-state energy difference between \isotope[24]{O} and \isotope[28]{O}.
    The training and validation data are given with the blue and orange circles, respectively.
    }
    \label{fig:em}
\end{figure*}
The VS-IMSRG has been one of the very successful many-body techniques in nuclear theory.
To make a robust prediction through the VS-IMSRG calculations, one would need to run the million or billion times calculation, which is clearly unrealistic.
Therefore, the construction of a fast and accurate VS-IMSRG emulator is an essential task.
The widely applied EC method, however, does not allow us to construct the emulator, as access to the VS-IMSRG eigenvector is not straightforward.
Parametric Matrix Model (PMM) is a class of newly developed model- and data-driven hybrid emulators~\cite{Cook:2024toj}. 
In contrast to the EC method, the PMM needs only the eigenvalues, making the emulator construction feasible.
We have observed that the PMM often performed better than more general data-driven models, including the Gaussian Process on smaller datasets~\cite{Somasundaram:2024zse}.

We briefly discuss PMM for the VS-IMSRG and validate its performance.
For emulators built directly from LECs input $c_i$, we solve the following equation:
\begin{equation}
    (H_0 + c_1 H_1 + c_2 H_2 +\ldots + c_n H_n - E I ) \mathbf{y} = 0\,,
    \label{eq:em-PMM}
\end{equation}
It is almost the replication of Eq.~\eqref{eq:emu}, with the choice of norm kernels to be always $I$. 
The matrix elements of $H_i$ are acquired by a machine learning method that minimizes a loss function with the predictions in the following steps: First, we generate random hermitian $H_i$. 
Then we compute the (usually lowest) eigenvalue of Eq.~\eqref{eq:em-PMM} $E_{\rm PMM}$ with a given training set $\{c_i\}$. 
The difference between training target $E$ and $E_{\rm PMM}$ will enter the loss function.
Finally, we use (Adam~\cite{kingma:2017ada}) gradient descent on the $H_{i}$ elements to optimize the loss function, hence the name ``Parametric Matrix''. 
In our case, we choose the simple mean absolute loss function $l_{\rm{MAE}}$ that averages the absolute difference between the PMM and the VS-IMSRG results:
\begin{equation}
    l_{\rm{MAE}}[\{H_{i}\}_{i=0, \ldots, n}] = \frac{1}{N_{\rm train}} \sum_{k=1}^{N_{\rm train}} |E_{k} - E_{k, {\rm PMM}}|\,.
\end{equation}

We estimate the required number of training samples $N_{\rm train}$ with the \isotope[16]{O} data (from Subspace-Projected Couple Cluster method~\cite{Ekstrom:2019lss}) provided in the NI dataset\footnote{This dataset originally consisted of 8,188 data points, later updated to 8,218; the additional 30 points are unlikely to alter our results.}.
As illustrated in Fig.~\ref{fig:em_PMM}, it is seen that our PMM emulator with the 17 active LECs trained with the randomly selected 300 NI samples works well. 
We observed that $N_{\rm train} = 300-500$ with $12 \times 12$ PMM matrix size provides the accurate results for the total 8188 NI samples with the mean absolute error of 0.1 MeV, which is smaller than other theoretical uncertainties~\cite{Hu:2021trw, Kondo:2023lty}. 


This estimation indeed holds when emulating VS-IMSRG calculations of \isotope[24]{O} in Fig.~\ref{fig:em_24O}.
Moreover, Fig.~\ref{fig:em_MEPMM} demonstrates a good performance of our MEP emulator, trained with the 17 input observables, i.e., the NN scattering phase shifts and ground-state energies of the few-body systems.
In these figures, with only 300 samples available, we rely on jackknife analyses to test the general performance of our method.
We randomly choose a subset of 250 data points from the 300 samples. Then, for each jackkife set, we take 240 random resamples from this subset and train a new MEP emulator. 
The whole training takes roughly 10 minutes for 100 batches on a laptop computer. Finally, the averages of these 100 emulators and predictions on the remaining 50 data points are displayed in our Figs.~\ref{fig:em_24O}~\ref{fig:em_MEPMM}~\ref{fig:em_MEPdiff}.

As one can easily find that the affine form is preserved for an energy difference in the MEP emulator, we can take $E_{\isotope[24]{O}} - E_{\isotope[28]{O}}$ as our output:
\begin{multline}
    \Big[\mathcal K _{i0} + E^{[1]} \mathcal K _{i1}  + \ldots+ E^{[n-1]} \mathcal K _{i(n-1)} + E_{\isotope[24]{O}}(\mathcal K _{in} - \mathcal K _{0})\\ +
    (E_{\isotope[24]{O}}-E_{\isotope[28]{O}})\mathcal K _{0}\Big]\mathbf{y} = 0.
    \label{eq:em_MEPdiff}
\end{multline}
Eq.~\eqref{eq:em_MEPdiff} allows us to emulate a differential quantity directly, which incorporates a cancellation of the systematic uncertainty of the theoretical calculations.
We then present our final validation in Fig.~\ref{fig:em_MEPdiff}. 
This is a pure MEP prediction with training data generated from the VS-IMSRG calculations described in the main text.
We find the mean absolute error from our resampling to be 0.2 MeV. 
This would be acceptable in the current application because the basis truncation error is larger, i.e., the energy difference between $e_{\rm max}=12$ and $16$ results is about $0.6$ MeV for \isotope[28]{O}. 
This bias will shift $E_{\isotope[27]{O}} - E_{\isotope[28]{O}}$ by $+0.2$ MeV. Basis truncation errors are closely related to the continuum effect~\cite{Luscher:1990ux}. The previous estimation of the effect on energy differences $E_{\isotope[28]{O}} - E_{\isotope[24]{O}}$ owing to continuum is around 1 MeV \cite{Hagen:2016xjv}. 
Including these continuum effects and basis truncation errors in the VS-IMSRG is beyond the scope of this work.